\documentclass[12pt]{article}
\usepackage{amsmath}
\usepackage{amssymb}
\newcommand{\be}{\begin{equation}}
\newcommand{\ee}{\end{equation}}
\newcommand{\lan}{\langle}
\newcommand{\rrr}{\rangle}

\usepackage{graphicx}
\usepackage{dcolumn}
\usepackage{bm}

\begin{document}

\begin{titlepage}

\begin{center}

{\Large Finite time measurements by Unruh-DeWitt detector and Landauer's principle}

\bigskip

{V.Shevchenko}

\bigskip

{\it National Research Centre "Kurchatov Institute" \\
ac.Kurchatova sq., 1, Moscow 123182 Russia}

\medskip

{\it Far Eastern Federal University, \\
Sukhanova str. 8, Vladivostok 690950
Russia}

\bigskip

\end{center}

The model of Unruh-DeWitt detector coupled to the scalar field for finite time is studied. A systematic way of computing finite time corrections in various cases is suggested and nonperturbative effects like thermalization are discussed. It is shown in particular that adiabatic switching off the coupling between the detector and the thermal bath leaves non-vanishing corrections to the detector's levels distribution. Considering the two level detector as an information bearing degree of freedom encoding one bit of information, limits on external work for the detector's (de)couling in finite time following from the Landauer's bound are formulated.

\end{titlepage}

\section{Introduction}

The problem of measurement occupies a special place among cornerstones of quantum theory \cite{braginsky-1,m,all}.
The measurement procedure, in general case, refers to at least two physical systems (or two distinct parts of a system), interacting with each other, one denoted as "the system" being measured, while the other is "the detector". There is a crucial assumption about one's ability to fully control the detector dynamics, in other words, the properties of the detector subsystem should be perfectly understood (and lack of such understanding inevitably brings systematic error to the results). Another important point is that the detector subsystem state at some chosen initial moment is supposed to be known, i.e. being read by another detector (our brain, for example) with certainty. Then, reading the detector state ("the measurement result") at some later moment, one can come to conclusions about properties of the system the detector has been interacting with. The details vary in wide range - in particular, impossibility (counterintuitive) to perform a measurement with negligible influence on the system is a crucial feature of the quantum world.

The problem of measurement in quantum field theory is even more complicated than in quantum mechanics due to the fact that the number of degrees of freedom is not fixed in the former case \cite{lp}. Using path integral language, one defines typical theory in terms of action $S[\phi]$ and integration measure ${\cal D} \phi$, and dynamical content of the theory is usually assumed to be encoded in the action and not in the measure. But in fact this is a matter of choice and the dynamics can be "redistributed" between measure and action in arbitrary proportion.\footnote{In abelian vector field theory, for example, instead of integration over vector field ${\cal D} {A_\mu}$ one can equivalently integrate over tensor fields with the Bianchi identities constraint ${\cal D} F_{\mu\nu}\> \delta(\partial\tilde{F})$. The standard $F^2$-action contains derivatives and describes propagation for the former measure but is trivial Gaussian local one with the latter measure.} Moreover, the integration measure can encode some {\it a priopri} known (or assumed) results of measurements, for example, boundary conditions on the fields \cite{bordag}.

Another instructive example is UV-regularization, corresponding to limiting the integral over fields $\phi(k)$ to the domain of momenta ${|k|< \Lambda}$, which in this language is nothing but some assumption about the fields beyond this boundary (surface in momentum space). In renormalizable theories all physics above $\Lambda$ can be encoded just in a couple of numbers - coefficients in front of marginal operators like the famous fine structure constant $1/137$ in QED. Renormalizability, however, does not mean naturalness. Computing, for example, average of some local $d$-dimensional operator ${\cal T}[\phi]$ in a theory with UV cutoff $\Lambda$, one gets typically (up to, perhaps, logarithms):
\be
\int {\cal D} \phi \>   {\cal T}[\phi] \>  e^{iS[\phi]} \sim  c\cdot \Lambda^{d} + {\mbox{finite part}}
\ee
The first term characterizes {\it ad hoc} assumptions about the integration measure/detector (for example, geometry of the lattice with the link size $a\sim 1/\Lambda$ used for computation), but not the genuine physics of this operator.
Of course in many cases symmetries of the theory guarantee $c=0$, but if not, we have to work out a way of disentangling "the physics of the detector" from "the physics of the physics" (which presumably is hidden in the finite part). Notable examples include vacuum energy density, Higgs boson mass and gluon condensate, where only in the latter case the answer is known (qualitatively, but not quantitatively) - quantum scale anomaly of Yang-Mills theory and dimensional transmutation take care of unphysical UV scale $\Lambda$ to become physical scale $\Lambda_{QCD}$. In other cases of this sort, like cosmological constant problem or hierarchy problem a solution is yet to be found.

But in most cases in particle physics we assume that dynamics, described by the action is uncorrelated with dynamics of the measure and for good reasons: typical scale of the former is given by strong interaction distance of $\sim 10^{-15}$ meters and even smaller for weak interactions, while detectors are macroscopic objects having sizes of the order of dozens of microns and larger. All the standard perturbative quantum field theory machinery (asymptotic states; propagators computed in plane waves basis etc.) is based on this assumption. Even if the detector dynamics is relevant, like in case of Unruh effect and similar phenomena - one usually tries to disentangle "beautiful" field theoretic part (universal response functions etc) from "ugly" detector part (concrete models of the detector). On the other hand, to what extent it is possible is undoubtedly quantitative question which should be analyzed in each particular problem.

The measurement problem has been given another interesting prospective by introducing the concept of information. The subject's roots go back to the XIX-th century. Seminal insights of J.C.Maxwell, L.Boltzmann, J.von Neumann, L.Szilard, L.N.Brillouin and many others shaped this area of research. In modern science, the work of Claude Shannon \cite{cs} put the intuitive idea of information on firm mathematical grounds.
It happened to be the closest relative of entropy, quantifying delicate relations between micro- and macrostructures of various systems. Moreover, it was demonstrated long time ago \cite{jaynes} how to get all the main results of equilibrium thermodynamics from the information theory (for more recent exposition see \cite{fw}). Nevertheless, despite tremendous developments, it still and rightfully motivates many researchers to study various aspects of nontrivial interplay between purely combinatorial (and geometrical) and dynamical facets of the information/entropy.

There is another crucial point here. The physical information must be linked to real systems used to produce, to store, to transmit and to erase it. All information processing systems ("hardware") of today or tomorrow are to obey the laws of physics, but much less trivial is a question about intrinsic energy or entropy cost of "software" - algorithms of computation. Are there any fundamental limits from this point of view? A typical question of this kind looks like: does the amount of energy one has to spend to perform some operation with information depend on the nature of this information and if yes, how? In particular, can one imagine a device capable to copy one bit at zero energy cost? Is there minimal time needed to copy (or to erase), etc? Such problems are under discussion for decades (see \cite{bek,hsu,bek-2,rl1}) and are of prime importance for information theory understood as the one among other natural sciences. Of course, answers to these questions for real computers of today could be very different quantitatively from the ones for idealized devices.

It is easy to understand that in general one can indeed expect various limits of this kind. For example, limiting character of the velocity of light puts relativistic bounds on the computing speed of any device having finite size (as should be the case for any realistic one). Speed of quantum evolution is limited by the so called Margolus-Levitin bound \cite{ml}, inclusion of gravity puts its own limits, and so on \cite{bek,lloyd,bremer,rl3}.

In his seminal paper \cite{rl}, Rolf Landauer made an important contribution to this discussion. He formulated the statement known as the Landauer's principle: erasure of one bit of information by a device at temperature $T$ leads to dissipation of at least $k_B T \log 2$ of energy. Roughly speaking, forgetting is costly.\footnote{As is nicely said in \cite{ball}, - "Even if you're not burning books, destroying information generates heat".}
By "erasure" one understands any (almost) operation that does not have single-valued inverse. One the other hand, reversible  logical operations such as copying can be performed, at least in principle, with zero energy consumption, i.e. for any chosen $\epsilon > 0$ one can suggest copying protocol which requires an amount of energy less that $\epsilon$.

Landauer's principle is considered by many as a key to solution of the famous Maxwell demon paradox (see \cite{jn,adami,bennet,bennet2,earman,vedral-ph} and nice introductory review to the subject \cite{plenio}). The crucial point is the necessity to restore initial state of the demon's mind ("reset its memory"), and this operation dissipates just the amount of heat one gained during the previous steps of the demon's activity. The discussions between different schools of the Maxwell demon's exorcists still go on.

It is rather challenging to explore this kind of physics experimentally and direct evidence of the Landauer's principle came quite recently \cite{ueda,ueda2,ara,goold2}. The validity of the principle was demonstrated, but two important comments are worth making. First, the bound is valid only statistically, while in a single event fluctuations can drive the system well below it. Second, the bound is saturated in stationary case, and dissipated heat increases with decrease of the erasure time.

More refined formulation of the principle was given recently in \cite{reeb}. The authors considered a system immersed into the thermal bath.
Initially, the density matrices of the system ("detector") and the bath are uncorrelated:
\be
\rho = \rho_s\otimes \rho_{b}
\
\ee
where $\rho_{b} \sim \exp(-\beta H_b)$ and $\beta$ is initial inverse temperature. After unitary evolution the density matrix $\rho'$ has no longer the form of direct product $\rho_s' \otimes \rho_{b}'$ and one defines $\rho_s' = \mbox{Tr}_b \rho'$, $\rho_b' = \mbox{Tr}_{s'} \rho'$. Then one can strictly prove that
\be
\beta \Delta Q \ge \Delta S
\label{l}
\ee
where $\Delta S = S(\rho_s) - S(\rho_s')$, $\Delta Q = \mbox{Tr}[H \rho_b'] - \mbox{Tr}[H \rho_b]$ with the standard definition of entropy $S[\rho] = - \mbox{Tr} [\rho \log \rho ]$.
The equation (\ref{l}) is a precise formulation of the stationary Landauer's principle. It is worth noting that factorization of the initial state density matrix and thermal character of the bath are important, without these assumptions one could design gedanken experiments violating (\ref{l}).

In its original form Landauer's principle corresponds to $\Delta Q > 0$, $\Delta S > 0$ case. The physical meaning is pretty clear - suppose we have a system in contact with a thermal bath, and observe that for whatever dynamical reason it gets more ordered than it was initially ($\Delta S > 0$, like it happens, for example, in crystallization process). As a limiting case one could think of total reset. It is a process with definite final state of the system regardless the initial state was and hence $S(\rho_s') = 0$. The Landauer's principle then states that the amount of thermal energy dissipated to the bath $\Delta Q $ is bound from below by (\ref{l}). For two level system encoding one bit of information the original Landauer's $ \log 2$ factor reappears.

The principle can be applied to other cases as well, being trivial in $\Delta Q > 0$, $\Delta S < 0$ case (this describes explosion-like processes when internal energy converts into heat) and equivalent to nonexistence of perpetuum mobile of the second kind for $\Delta Q < 0$, $\Delta S > 0$ case - energy of the thermal bath cannot be used to make external system more ordered. The fourth case $\Delta Q < 0$, $\Delta S < 0$ physically corresponds to using the thermal energy to disorder the system like it happens during melting.

The simplest possible case usually considered in the literature is two-state detector, encoding one bit of information.
 Two kinds of a detector design are being analysed. In the first case, two distinct states of a detector correspond to different space coordinates, like, for example, in the double-well experiments \cite{ara}. In the second case, the states are distinguished by energy coordinate, like, for example, the ground and excited states of an atom. One could think about another possibilities, even rather exotic like famous Schr\"odinger cat as the detector of cyanide presence in the box. It is remarkable that Landauer's principle pretends to be applicable in any of these cases, despite their physical dynamics is obviously far from being identical.

One can think also about different erasure protocols. The most trivial one is to reset the system, i.e. to put it by any means in some predefined final state (we label it as 0 in what follows). But other options are also possible. For example, the system can be thermalized \cite{lubkin} by long enough contact with the heat bath. Or, alternatively, one can shake up a detector by some strong external force, leaving it relaxed in some random final state.
In all cases any information about initial state of the detector is completely lost and this justifies the term "erasure", which should not be misinterpreted as "reset". The latter term corresponds only to the predefined final state of zero entropy.

One might be interested in the subject outlined above from a different prospective. Usually the reset operation is assumed to be done by some external agent, which, for example, removes the barrier between "0" and "1" states, deforms the potential well, etc. Then we are interested in the work done by this external force. On the other hand, one can imagine the detector having some internal time scale to be reset. Good example is an excited atom, where its decay time plays the role of this scale. In this case no external work is needed. Since energy $\Omega$ emitted by the atom (detector) and swallowed by the bath can be smaller than $k_B T \log 2$, naively (\ref{l}) looks to be violated. This is of course not the case since the total reset is incompatible with final thermal equilibrium between the detector and the bath - there is always exponentially suppressed at low temperatures but nonzero probability to find the detector in its excited state, and the corresponding correction to the right hand side of (\ref{l}) exactly compensates the bath energy deficit in the left hand side mentioned above. But it is by far not obvious whether this will also be true for finite time interval, i.e. before equilibration.

 In the present paper we develop formalism for systematic studies of Unruh-DeWitt point-like detector coupled to the field bath for finite time. Various aspects of this problem have been discussed in the literature \cite{padm,padm2,barbado,satz} and the work presented here is heavily based on these findings. The main focus of the present paper is to address this sort of questions in a way, which is suitable not only in the leading order in perturbation theory (where most work has been done so far) but also beyond it. As an application, we discuss limits on external work for the detector's (de)couling in finite time, which are put by the Landauer's principle, if the detector is understood as an information bearing degree of freedom encoding one bit of information.

\section{Theory of finite time measurements}

\subsection{Basic definitions}

In this section we discuss general formalism of Unruh-DeWitt point-like detector model \cite{unruh,dw,ta,birrel} for finite-time measurements, corresponding to energy coordinate, distinguishing the states. The infinite time limit of this model is well studied. The case with explicit time-dependence attracted less attention, however many interesting results are obtained here (see, e.g. \cite{padm,padm2,barbado,satz} and recent papers \cite{kempf,garay}). The detector-field system state is described as a vector from the space $|n, \Phi \rrr $, where the index $n$ encodes discrete state of the detector, while $\Phi$ stays for state of the field subsystem. For two-level Unruh-DeWitt detector immersed in the heat bath the index $n$ takes values 0 or 1 and $\Phi_\beta$ stays for thermal state of free massless scalar field with inverse temperature  $\beta = (k_B T)^{-1}$.

The evolution operator we focuss our attention on in this paper reads:
\be
U_\chi = {\mbox{T}}\exp\left[ ig\int d\tau \> \chi(\tau) {\bf\mu}(\tau)  \phi(x(\tau)) \right]
\label{ude}
\ee
Here $x(\tau)$ parameterizes the detector's world-line, $\tau$ is a proper time along it, ${\bf\mu}(\tau)$ - monopole transition operator for the detector and the latter is assumed to evolve with Hamiltonian $H_d$ having discrete spectrum $\{ E_n \}$ . The field-detector coupling constant is $g$ and the field $\phi(x(\tau))$ is assumed to be elementary massless scalar field. Operators are ordered along the world-line by the standard T-ordering recipe. If the detector is at rest, the proper time can be chosen to coincide with the usual time, $x(\tau) = (\tau, 0, 0, 0)$.

The important ingredient is the window function $\chi(\tau)$, parameterizing the measurement procedure. In equivalent way one can speak about $g\chi(\tau)$ as about time-dependent coupling, which is assumed to be different from zero in a given proper time intervals.
In what follows we consider two different kinds of the window function (real and non-negative for any $\tau$ in all cases). The first is "switch off" function $\chi_-(\tau)$:
\be
\chi_-(\tau) \to 1  \mbox{ for }  \tau \ll \bar\tau  \;\;\; ; \;\;\; \chi_-(\tau) \to 0  \mbox{ for }  \tau \gg \bar\tau
\ee
We will speak about $\bar\tau$ as about the moment of switching the detector off.
Analogously, "switch on" function $\chi_+(\tau)$ is given by $\chi_+(\tau) = 1 - \chi_-(\tau)$.

The second type is a finite window function $\chi(\tau)$ with the property:
\be
\chi(\tau) \sim 1  \mbox{ for }  |\tau - \bar\tau|/\tau_m \lesssim 1 \;\;\; ; \;\;\; \chi(\tau) \to 0  \mbox{ for }  |\tau - \bar\tau|/\tau_m \to \infty
\ee
Here $\bar\tau$ is natural to identify with {\it the moment} of the measurement, while $\tau_m$ is the measurement's {\it duration}. The simplest case (sharp boundaries) is given by
\be
\chi_{12}(\tau) = \theta(\tau - \tau_1) - \theta(\tau - \tau_2)
\label{olk}
\ee
where we take $\tau_2 > \tau_1$ and the standard $\theta$-function is $1$ for $x>1$ and $0$ for $x<0$. The operator (\ref{ude}) with the function $\chi_{12}(\tau)$ describes evolution from the moment $\tau_1$ till the moment $\tau_2$:
\be
U_{\chi_{12}} = U(\tau_1, \tau_2) = {\mbox{T}}\exp\left[ ig\int\limits_{\tau_1}^{\tau_2} d\tau \> {\bf\mu}(\tau)  \phi(x(\tau)) \right]
\label{ude2}
\ee
This function obeys the standard composition law
\be
U(\tau_1, \tau_2)U(\tau_2, \tau_3) = U(\tau_1, \tau_3)
\label{kij}
\ee
with obvious consequence $U(\tau_1, \tau_2)U(\tau_2, \tau_1) = 1$. Infinite time (stationary) measurement corresponds to the limits $\tau_1 \to -\infty$, $\tau_2 \to \infty$.

It is useful to introduce evolution for semi-infinite interval:
\be
U_\chi(t) = {\mbox{T}}\exp\left[ ig\int\limits_{-\infty}^t d\tau \> \chi(\tau) {\bf\mu}(\tau)  \phi(x(\tau)) \right]
\label{si}
\ee
The amplitude for the detector's transition to the state $|n\rrr $ from another state $|m \rrr$ is given by the following expression:
\be
{\cal A}_{m\to n} = \lan \Phi', n | U_\chi |m, \Phi \rrr
\label{aex}
\ee
The state $| \Phi' \rrr$ represents final (after the detector's transition) state of the field subsystem, while initially it is supposed to be in the initial state $|\Phi \rrr$. Transition probability summed over final states of the field subsystem reads:
\be
{\cal P}_{m\to n} = \sum_{\Phi'} \left| {\cal A}_{m\to n} \right|^2
\label{pex}
\ee
and unitarity dictates $\sum_n {\cal P}_{m\to n} = 1$. For two-level detector it reads
\be
{\cal P}_{0\to 0}(t) = \lan 0,\Phi | U_\chi^\dagger(t) |0\rrr \lan 0 | U_\chi(t) |0, \Phi \rrr
\label{p0}
\ee
where the sum over intermediate states of the field subsystem has been already taken and due to completeness relation $|0\rrr \lan 0 | + |1\rrr \lan 1 | = 1$ the probability is normalized
\be
{\cal P}_{0\to 0}(t) + {\cal P}_{0\to 1}(t) = 1
\label{n2}
\ee
as it should be.

It is well known \cite{padm,satz,kempf} that the detector evolution can be sensitive to whether switching functions $\chi(t)$ are smooth or sharp.
We consider only smooth functions $\chi(t)$ with smooth derivatives $\chi'(t)$ in this paper, and discuss sharp boundaries only as a limiting case.

It is clear that detailed profile of the resulting probability depends on the profile of the function $\chi(t)$. The situation greatly simplifies in weak coupling regime. Let us remind the traditional treatment when probability is computed in perturbation theory. Formal expansion over $g$ gives expressions at the leading order
\be
{A}_{m\to n} = i g\int d\tau \> \chi(\tau) \lan n | {\bf\mu}(\tau) | m \rrr \cdot \lan \Phi' | \phi(x(\tau))| \Phi \rrr
\ee
and
\be
P_{m\to n} =  {\bar g}_{mn} \>\int d\tau \int d\tau' \> \chi(\tau) \chi(\tau')  \> e^{-i\Omega_{mn}s} \> G^+(\tau, \tau')
\label{p8}
\ee
where $s = \tau - \tau'$, ${\bar g}_{mn} = g^2 | \lan n | {\bf\mu}(0) | m \rrr |^2$, $\Omega_{mn} = E_n - E_m$ and the Wightman (positive frequency) function is given by
\be
G^+(\tau, \tau') = \lan \Phi | \phi (x(\tau)) \phi (x(\tau')) | \Phi \rrr
\label{wi2}
\ee

In case of infinite measurement $ \chi(\tau)\equiv 1$ the expression (\ref{p8}) diverges.
Typical approach is to consider transition probability in unit time instead, given by the ratio
\be
p_{m\to n} = \frac{P_{m\to n} = \infty }{\mbox{time} = \infty} \neq \infty
\label{psmall}
\ee
Naively generalization of (\ref{psmall}) to finite time case would look like $ P_{m \to n} = \tau_{m} \cdot p_{m\to n}$ where $\tau_{m}$ is the measurement time. This expression however is only a part of the right result, as we discuss in details below.
Full evolution picture for the probability is to be used to get the correct answer.

\subsection{Evolution equations}

General physical reasoning tells that for the detector and the bath being in contact for sufficiently long time, equilibrium should be reached with the detector's levels occupied according to the thermal distribution with the bath's temperature. Despite this final answer is clear, to show that explicitly could be tricky for a particular detector-bath coupling model. One possible way is to check that the transition rate of a stationary detector in a Kubo-Martin-Schwinger state satisfies the detailed balance condition (see detailed discussion in this framework in \cite{ff9}). Another approach is to study time evolution of the detector directly. For the monopole coupling (\ref{ude}) the most straightforward way to do that known to the author is presented in \cite{nitzan} (see also the paper \cite{brenna} where thermalization aspects are discussed and recent review  \cite{br} with references therein). For a reader's convenience we expose in Appendix A the main steps from that paper. The result (in approximations, discussed in the Appendix A) has the following form:
\be
\frac{d {\cal P}_{0\to 0}(t)}{dt} = - (C_+(t) + C_-(t) ) \cdot {\cal P}_{0\to 0}(t) + C_+(t)
\label{pp24}
\ee
and
\be
\frac{d {\cal P}_{0\to 1}(t)}{dt} = - (C_+(t) + C_-(t) ) \cdot {\cal P}_{0\to 1}(t) + C_-(t)
\label{pp214}
\ee
where $C_+(t)$ is coefficient function $C(t, \Omega) $ given (in unregularized form) by
\be
C_+(t) \equiv C(t, \Omega)  =
{\bar g}\int\limits_{-\infty}^t d \tau \chi(t) \chi(\tau)  \left( G^+(t,\tau) e^{i\Omega (t-\tau)} + G^+(\tau,t) e^{-i\Omega (t-\tau)}\right)
 \label{gb1}
 \ee
and $ G^+(t,\tau)$ is the Wightmann function (\ref{wi2}). Analogously $C_-(t)$ stays for $C(t,-\Omega)$ and we denote ${\bar g} = g^2 \mu_0^2$ for brevity. It is worth noticing that $C_+(t) + C_-(t)$ is even function of $\Omega$ while each summand separately is not. We denote the energy levels of the detector as $E_0$ and $E_1$, having, by definition, $\Omega = E_1 - E_0 \ge 0$.

It is easy to see that equation (\ref{pp214}) is nothing than the standard  Markovian evolution with time-dependent coefficients,
\be
\dot{p} = - p\cdot p_{1\to 0} + (1-p)\cdot p_{0\to 1}
\label{jh}
\ee
where $p_{m\to n}$ are transition probabilities in unit time and the detector is described by the density matrix $
\rho = p \> |1\rrr \lan 1 | + (1-p) \> |0 \rrr \lan 0 |
$, then the detector's energy is given by $
E_d = pE_1 + (1-p)E_0 = p\Omega + E_0 $
and the corresponding entropy $
S_d(p) = - p\log p - (1-p) \log (1-p) $.
General solution for dynamics of the detector is given by
\be
{p}(t) = p(t_0) \cdot {\cal P}_{1\to 1}(t_0, t)  + (1- p(t_0)) \cdot {\cal P}_{0\to 1}(t_0, t)
\label{o2}
\ee
or $ {p}(t) = {\cal P}_{0\to 1}  + p(t_0) \cdot (1 - {\cal P}_{1\to 0}  - {\cal P}_{0\to 1}) $
 where we omitted time arguments in transition probabilities for simplicity of notation and took into account normalization of nonnegative transition probabilities (\ref{n2}) in finite time.
The dissipative dynamics corresponds to losing information about initial occupancy $p(t_0)$ (in other words, $ {\cal P}_{1\to 1} $ becomes indistinguishable from $ {\cal P}_{0\to 1} $ for large times).



To proceed further we are to regularize and renormalize (\ref{gb1}) since formally the coefficient functions are UV-divergent. Notice that, as is usual in quantum field theory, the first step of this procedure (regularization)
 is formal and can be realized in many alternative ways, while the second one (renormalization) is to be based on physically motivated requirements put on renormalized quantities $C^{(r)}_\pm(t)$.

The first step is regularization of the Wightman function (\ref{wi2}). In general, this procedure is rather tricky, see  \cite{schlicht,louko,satz,louko2,barbado} where various aspects of it are discussed. We do not repeat the arguments from these papers here referring the interested reader directly to the original publications. In summary, the physical meaning behind reflects an obvious fact that there are no point-like detectors and any realistic detector has finite size and therefore some intrinsic finite internal time given by this size divided by the speed of light. This time plays a role of UV-regulator.

We are considering standard thermal Wightman function in this paper
\be
G^+_\beta(\tau, \tau') = - \frac{1}{4\pi^2} \sum\limits_{l=-\infty}^{\infty} \frac{1}{(\tau - \tau' + i\beta l)^2} = - \frac{1}{4\beta^2} \frac{1}{\sinh^2\left(\pi(\tau - \tau' )/\beta\right)}
\label{wight}
\ee
Only the free part, corresponding to $l=0$ term, leads to UV-divergencies, and following the work mentioned above we regularize it as:
\be
 \frac{1}{(\tau - \tau' )^2} \rightarrow   \frac{1}{(\tau - \tau' - i\epsilon\zeta(\tau, \tau'))^2}
\ee
where the infinitesimal regulator parameter $\epsilon > 0$ physically represents the finite size (divided by $c$) of a realistic detector. The function $\zeta(\tau, \tau')$ can be taken as unity in the simplest case of the detector at rest, however in general case $\tau$-dependence of $\zeta(\tau, \tau')$ is important to get correct Lorentz-invariant and causal answers (see discussion in \cite{schlicht}, where $\zeta(\tau, \tau') = \dot{x}(\tau) + \dot{x}(\tau')$
prescription is advocated).

Taking into account that $ i(G^+_\beta(t, \tau) - G^+_\beta(\tau, t)) \xrightarrow[\epsilon\to 0]{\phantom{\epsilon\to 0}} - \frac{1}{2 \pi} \delta'(s) $
where $s = t -\tau$, and $\delta'(s)$ is delta function derivative, one can rewrite (\ref{gb1}) as
\be
 C(t, \Omega) = {\bar g}\chi(t) \int\limits_{0}^\infty d s \> \chi(t-s)  {\tilde G}^+_\beta(s,t) \cos\Omega s + {\bar g} \frac{\Omega}{4 \pi} \chi^2(t)
 \label{gb2}
 \ee
where $ {\tilde G}_\beta^+(s,t) =  G_\beta^+(t,t-s) +  G_\beta^+(t-s, t)$ and for the chosen regularization with $\zeta(\tau, \tau') \equiv 1$ the second argument can be omitted, $ {\tilde G}_\beta^+(s,t) = {\tilde G}_\beta^+(s)$.

\subsection{Infinite time measurements}

Let us consider first the $\chi(\tau)\equiv 1$ case. Then, physical condition for renormalization should be $C_-(t) = C(t, -\Omega) \equiv 0$ for $T=0$, which means, that the detector at rest located in the vacuum at zero temperature cannot be excited and ${\cal P}_{0\to 1}(t) \equiv 0$.
It is easy to see that this condition is satisfied by the chosen regularization since
\be
\int\limits_{0}^\infty d s \> {\tilde G}_\beta^+(s) \cos\Omega s = \frac12 \left( F_\beta(\Omega, \epsilon) + F_\beta(-\Omega, \epsilon)\right)
\label{30}
\ee
where
\be
F_\beta(\Omega, \epsilon) = \int\limits_{-\infty}^{\infty} ds \> e^{-i\Omega s} \> G^+_\beta(s) =
\frac{\Omega}{2\pi} \>\frac{e^{\epsilon\Omega} }{e^{\beta\Omega} - 1}
\label{f1}
\ee
In zero temperature limit
\be
F_\beta(\Omega, 0)\to F_\infty(\Omega, 0) = -\Theta(-\Omega) \> \frac{\Omega}{2 \pi}
\ee
and indeed $C(t,-\Omega) = 0$ (let us remind that $\Omega > 0$).
There is important relation between the functions $F_\beta(\Omega, \epsilon)$ and $F_\beta(-\Omega, \epsilon)$
\be
F_\beta(-\Omega, \epsilon) - F_\beta(\Omega, \epsilon) = \frac{\Omega}{2\pi} - 2\sinh\frac{\epsilon \Omega}{2} \left[ F_\beta(\Omega, \epsilon/2) + F_\beta(-\Omega, \epsilon/2) \right]
\label{ff}
\ee
where the second term in the right hand side of (\ref{ff}) vanishes in the limit $\epsilon \to 0$. It is worth noticing that $\Omega$ is kept finite when $\epsilon \to 0$. To simplify notation we drop the second argument of $F_\beta(\Omega, \epsilon)$ and denote $F_\beta(\Omega)=F_\beta(\Omega, 0)$ in what follows. Notice also that
\be
F_\beta(\Omega) + F_\beta(-\Omega) = \frac{\Omega}{2\pi} \frac{e^{\beta\Omega} + 1}{e^{\beta\Omega} - 1}
\ee
is, of course, even function of $\Omega$, but its formal zero temperature limit $\Omega/2\pi$ is odd.

If the detector is initially in the ground state ($p(t_0)=0$), solving equation (\ref{pp214}) one gets
\be
{p}(t) = \int\limits_{-\infty}^{t} d\tau \> C_-(\tau)\>  e^{-{\int\limits_{\tau}^{t} (C_+(\tau') + C_-(\tau')) d\tau'}}
\label{p01}
\ee
One can easily check that in infinite measurement case $C_\pm(t)$ are $t$-independent factors given by
\be
C_-(t) =  {\bar g} \int\limits_0^{\infty} ds \left( G_\beta^+(s) e^{-i\Omega s} + G_\beta^+(-s) e^{i\Omega s}\right) =  {\bar g} F_\beta(\Omega)
\label{p03}
\ee
and $C_+(t) = {\bar g} F_\beta(-\Omega)$, so that
\be
{p}(t) = {\bar g} F_\beta(\Omega) \int\limits_{-\infty}^{t} d\tau \> e^{-{\bar g} (t-\tau) (F_\beta(\Omega) + F_\beta(-\Omega) )}
\label{p06}
\ee
which is of course $(1+e^{\beta\Omega})^{-1}$ as it should be. Notice that the same result one would get for $\chi(\tau) = \theta(t-\tau)$ (abrupt switching off at the moment $t$) instead of $\chi(\tau) \equiv 1$ (no switching off at all).

Let us make one more general remark. To get thermalization in infinite measurement case one has no need to use dynamical solution (\ref{p01}), instead the stationary point of (\ref{pp24}), (\ref{pp214}) can be considered:
\be
{{\cal P}}_{0\to 0} = - \lim\limits_{t\to\infty} \> \frac{C_+(t)}{C_+(t) + C_-(t)}
\label{pp5}
\ee
and plugging (\ref{f1}) and (\ref{p03}) into (\ref{pp5}) one gets expected answer
\be
{{\cal P}}_{0\to 0} = \frac{F_\beta(- \Omega)}{F_\beta(\Omega) + F_\beta(-\Omega)} = \frac{1}{1+e^{-\beta\Omega}}
\label{prob}
\ee
corresponding to the full thermalization of the detector. The same result can be obtained, of course, for probability ${{\cal P}}_{1\to 0} $,
which is equal to ${\cal P}_{0\to 1} = 1-{\cal P}_{0\to 0}$ with the change $\Omega \to - \Omega$. For infinite time interaction the detector's probability to occupy a given final state does not depend on its initial state.

The remarkable property of the result (\ref{p06}), (\ref{prob}) is their independence on the coupling parameter ${\bar g}$, caused by the fact that the detector is on for infinite time and its excitations and de-excitations (whose balance produces (\ref{prob})) are controlled by the same coupling constant $g$. In other words,  with $\chi(t)\equiv 1$ the coupling in the model (\ref{ude}) defines internal clock rate via redefinition $g d \tau \to d \tau$ which is irrelevant in infinite integration time limit.

\subsection{Finite time measurements: renormalization formalism}

Let us now come to the case of non-stationary measurements. There can be various forms of time-dependence. Most work has been concentrated on effects resulting from the choice of the detector's trajectory $x(\tau)$. The best known result in this respect is celebrated Unruh effect itself \cite{unruh}, which established the correspondence between two response functions computed in the leading order in $g$ for infinite time measurement: one for $x(\tau)$ describing moving detector with constant
acceleration $A$ and interacting with free massless scalar field $\phi(x(\tau))$ at zero temperature; and another for $x(\tau)$ describing
detector at rest in the thermal bath of free massless scalar field $\phi(x(\tau))$ at temperature $T= \hbar A / 2\pi k_B c$. This seminal result is 40 years old, nevertheless it is still being intensively discussed (see, e.g. \cite{a}).

Much less attention has been paid to results not relying on the lowest order approximation in the coupling constant and for the finite measurement time case, corresponding to nontrivial function $\chi(\tau)$. It is sometimes said that non-universality of the results for $\chi(\tau) \neq 1$ makes the subject physically uninteresting since there is no clear way to separate the physics of the field $\phi(x)$ and of the measurement profile function $\chi(\tau)$ in this case. It is worth reminding in this respect that any actual measurement performed by a real physical device must always have finite duration. The question about influence of the measurement ("physics of $\chi(\tau) \neq 1$") on the system ("physics of $\phi$ ") must be answered for a particular problem in quantitative terms.

 There is another, even more fundamental reason, which makes finite time measurements interesting. The response function measures, roughly speaking, the field fluctuations pattern $\lan \phi \>\phi \rrr$ and not vacuum expectation value of the field itself $\lan \phi \rrr$ (see interesting discussion of this point in \cite{padms}). The latter however can be non-trivial classical solution to the field equations being constant or varying with its own space-time scale. In particular, it is obvious that measurement with finite $\tau_m$ is sensitive to the condensate $\lan \phi \rrr = $ const solution, while the infinite time measurement is not. The same is true with respect to other local vacuum averages. The physical reason is clear - to get the signal (i.e. the detector "click") some energy has to be transferred to it and this is impossible for the stationary detector being immersed not only into trivial vacuum but also in other, nontrivial ones, for example, with a broken symmetry. In other words, one has to give some time-dependent dynamics to the detector in order to be able to measure time-independent quantities (see, e.g. \cite{sh1,sh2,martin}).

Thus the main object of interest is probability to find the detector in excited state after the measurement is finished. In a complementary, ergodic fashion one can think about a macro-detector made of many elementary micro-detectors, each evolving according to (\ref{ude}). We are interested in the fraction of excited micro-detectors. After the interaction formally switches off, this fraction freezes and no more transitions are possible.

Generally speaking, there are at least four different time scales when we speak about non-stationary measurements. The first two characterize the detector - it is the detector energy levels difference $\Omega$ and regularization parameter $\epsilon$, related to the detector size. In general, they can be considered as unrelated ones, even parametrically. The second two parameters are external - it is the typical thermal time $\hbar/k_B T$ and speed of the coupling change $\chi'(\tau)/\chi(\tau)$. These two groups of parameters are related. For example, if $\chi'(t)\equiv 0$, there is no $\epsilon$ dependence. The outcome of non-stationary measurement, on the other hand, is in general $\epsilon$-dependent. Therefore, despite the fact that in principle the expression (\ref{p01}) can be evaluated numerically for any window function of interest, there is important intermediate step to be taken, related to renormalization of (\ref{gb2}), since this expression is formally divergent in $\epsilon \to 0 $ limit. The condition of vanishing of $C_-(t)$ at zero temperature
cannot be used anymore since switching process itself could excite the detector even in the vacuum.

To separate divergent contributions it is convenient to rewrite in (\ref{gb2}) $\chi(t-s) \to \chi(t-s)\pm \chi(t) \mp s\chi'(t)$ and
$  {\tilde G}_\beta^+(s) \to {\tilde G}_\beta^+(s) \pm {\tilde G}^+(s)$ where
\be
{\tilde G}_\beta^+(s) = - \frac{1}{4\beta^2} \left( \frac{1}{\sinh^2\left(\frac{\pi(s - i\epsilon )}{\beta} \right)} + \frac{1}{\sinh^2\left(\frac{\pi(s + i\epsilon)}{\beta} \right)} \right)
\ee
and zero temperature function (where we omit the $\beta$ index):
\be
{\tilde G}^+(s) = - \frac{1}{4\pi^2} \left( \frac{1}{(s - i\epsilon )^2}  + \frac{1}{(s + i\epsilon )^2} \right)
\ee
to get
\be
\begin{split}
 C(t, \Omega) = {\bar g}\chi^2(t) F_\beta(-\Omega) +  {\bar g}\chi(t)\chi'(t) \zeta + & \\
+ {\bar g}\chi(t) \int\limits_{0}^\infty d s \> (\chi(t-s) - \chi(t)) ( {\tilde G}_\beta^+(s)  - {\tilde G}^+(s) & (s)) \cos\Omega s + \\
+ {\bar g}\chi(t) \int\limits_{0}^\infty d s \> (\chi(t-s) - \chi(t) + s\chi'(t))  {\tilde G}^+(s&) \cos\Omega s
 \label{gb222}
\end{split}
\ee
where $ \zeta = - \int\limits_{0}^\infty d s \> s  \> {\tilde G}^+(s) \cos\Omega s $.

Let us discuss the structure of expression (\ref{gb222}). First of all, the expression (\ref{gb222}) is causal in a sense that $C_\pm(t)$ does depend only on $\chi(t)$ at the moments preceding $t$. The first term in the right hand side is generalization of (\ref{p03}) and it is the only contribution surviving in stationary limit. The third term is purely thermal contribution vanishing in $\beta \to \infty$ limit, while the fourth term is temperature-independent and describes detector's excitation/decay by switching procedure itself. Notice that these two terms are regular by construction, i.e. one can put $\epsilon = 0$ computing them. The second term is the only divergent (in $\epsilon \to 0$ limit) part of this expression. To compute make the structure of $\zeta$ more explicit, it is convenient to rewrite the integral for $\zeta$ as
 \be
 \zeta = \frac{1}{2\pi^2} \left( \int\limits_{0}^{\tau_s} ds +  \int\limits_{\tau_s}^{\infty} ds \right) \> s \>  \left( \frac{s^2 - \epsilon^2}{(s^2 + \epsilon^2)^2}\right) \cos\Omega s
  \label{po36}
  \ee
   where arbitrary positive parameter $\tau_s$ is assumed to be smaller than $1/\Omega$ but larger that $\epsilon $. Then
   \be
   \begin{split}
  \zeta =  & \frac{1}{2\pi^2}  \int\limits_{0}^{\tau_s} ds \>s\>   \frac{s^2 - \epsilon^2}{(s^2 + \epsilon^2)^2} + {\cal O} (\Omega^2 \tau_s^2 ) + \frac{1}{2\pi^2}  \int\limits_{\tau_s}^{\infty} ds  \>  \frac{\cos\Omega s}{s} + {\cal O} (\epsilon^2 / \tau_s^2 ) = \\
  = &  \frac{1}{2\pi^2}  \left( \log \frac{\tau_s}{\epsilon}  - 1  - \mbox{Ci}(|\Omega| \tau_s) \right) + ...
    \end{split}
  \label{spp}
  \ee
  The above expression is $\tau_s$-independent by construction. On the other hand, this parameter plays the same role as renormalization scale (subtraction point) in terms of conventional renormalization procedure. Indeed, it is clear from general energy-time uncertainty arguments that the probability for any realistic detector staying in its ground state to be excited should vanish in the limit $\Omega \to \infty$.
This is true for the second integral in (\ref{spp}) but not for the first one.
To get physical reason for this divergency it is useful to notice, first of all, that for our approach
to be valid, the detector-bath coupling should be approximately constant for the time of the order of $\epsilon$. If this is not the case, different parts of the detector start to interact nontrivially with themselves, resulting in possible detector's self-excitation. In other words, typical time $\chi(t) /\chi'(t)$ must at all $t$ be much larger than UV-cutoff $\epsilon$.
But this is not the end of the story. Any realistic detector has nonzero recovery time (in general, this parameter could be unrelated to $\epsilon$) and switching on and off at time scales smaller that this  time could not affect its counts. The parameter $\tau_s$ plays just this role in (\ref{spp}). In the formal limit $\tau_s \to 0$
\be
\zeta \to - \frac{1}{2\pi^2}  \left( \log \Omega \epsilon  + 1  + \gamma \right)
\ee

It is worth noticing that for switching functions vanishing at $\tau \to \pm \infty$ the integral $
 \int\limits_{-\infty}^{\infty} dt\> C(t, \Omega)
$ is regular and has no dependence on $\zeta$. In particular, this takes place at the leading order in perturbation theory (see below).
Formally speaking, this is due to the fact that integral $\int_{-\infty}^{\infty} d\tau \chi(\tau) \chi(\tau -s)$,
if convergent, is even function of $s$ and there is no linear term. Physically it reflects compensation between fragments of the detector's world-line where $\chi(t)$ is increasing and another ones where it is decreasing.
We can thus suggest the following prescription
\be
\zeta \to \zeta^{(r)} = - \frac{\mbox{Ci}(|\Omega| \tau_s)}{2\pi^2}
\label{z1}
\ee
for renormalized coefficients $C^{(r)}_\pm(t)$ (all other terms in (\ref{gb222}) are the same). This guarantees that $C^{(r)}_-(t) \to 0$ if $\Omega \to \infty$ provided $\tau_s$ is kept finite.

To summarize, the finite time measurement procedure we advocate in this paper should be augmented by  additional dimensionful parameter -
it is the detector size (in units of light-time) $\epsilon$ in unrenormalized language, or recovery time $\tau_s$ in terms of renormalized quantities. Both reflect micro-structure of the detector. Equations (\ref{pp24}), (\ref{pp214}) together with the renormalized coefficients $C^{(r)}_\pm(t)$ allow to compute numerically the resulting probability distribution for any given profile $\chi(\tau)$.

\subsection{Weak and strong coupling regimes}

It is instructive to distinguish weak and strong coupling regimes.
Physically the probability for the detector to be excited is small in weak coupling
limit and the leading contribution can be easily obtained from (\ref{p01})
\be
p = \int\limits_{-\infty}^{\infty} d\tau \> C_-(\tau)
\label{pert}
\ee
Since (\ref{pert}) is increasing with total measurement duration, it is clear that for large enough measurement time
one is always out of the perturbative domain. In the strong coupling limit, on the other hand, the detector "clicks" many times for the measurement time period. It means that for a general switching function the final answer is expected to be mostly controlled by its profile near the switch off moment and it does not depend on measurement time and on the details how the detector was switched on.

The convenient measure of perturbativity can be given by the speed of thermalization and hence initial state forgetting. It is related to the
difference
 \be
{\cal P}_{0 \to 0}(t) - {\cal P}_{1 \to 0}(t) = 1 - ({\cal P}_{0 \to 1}(t) + {\cal P}_{1 \to 0}(t)  )
\label{jh2}
 \ee
Taking into account (\ref{p01}) and analogous expression for ${\cal P}_{1 \to 0}(t)$, corresponding to $\Omega \to - \Omega$ change, one gets for this difference the following result:
 \be
 {\cal P}_{0 \to 0}(t) - {\cal P}_{1 \to 0}(t) = e^{-\int\limits_{-\infty}^t d\tau (C_+(\tau) + C_-(\tau))}
 \label{p12}
 \ee
 It is worth stressing that expression (\ref{p12}) is non-perturbative, i.e. it is valid for any value of the coupling constant.
 In $t\to \infty $ limit expression in the exponent takes the form
 \begin{equation}
 \int\limits_{-\infty}^\infty d\tau (C_+(\tau) + C_-(\tau))
=  {\bar g} \tau_m  \int\limits_{-\infty}^\infty d s \> D_\chi(s) ( G^+(s)  + G^+(-s) ) e^{i\Omega s}
 \end{equation}
where the function $D_\chi(s)$ is defined in \cite{ya2} (similar separation of variables was proposed in \cite{padm}):
 \be
 D_\chi(s) = \frac{1}{\tau_m}
  \int\limits_{-\infty}^\infty d\tau \> {\chi} \left(\tau + s/2\right) \chi\left(\tau - s/2\right)
 \label{d}
  \ee
and the total measurement time ${\tau}_m$ is defined as $
 \tau_m = \int\limits_{-\infty}^{\infty} d \tau \> (\chi(\tau))^2
 $.
Replacing in (\ref{d}) $s \to -i({\partial} / {\partial \Omega})$
the expression (\ref{p12}) can be written in terms of the operator $D_\chi(-\partial^2 / \partial\Omega^2)$:
\be
{\cal P}_{0 \to 0} - {\cal P}_{1 \to 0} = e^{-{\bar g} \tau_m ( D_\chi F_\beta(\Omega) + D_\chi F_\beta(-\Omega))}
\label{par}
\ee
Requirement for the dimensionless parameter in the exponent to be about unity defines critical time $\tau_m$  for transition of the problem from weak to strong coupling regimes.

An advantage of the expression (\ref{par}) is clear separation between the dynamics of the field subsystem, encoded in $F(\Omega)$, and the dynamics of measurement procedure, encoded in the operator (\ref{d}). This has been possible because of translation symmetry of the problem $G(\tau, \tau') = G(\tau-\tau')$, and can be generalized to other, non-thermal field ensembles. However a word of caution is to be said here. The operator (\ref{d}) is, generally speaking, a differential operator of infinite order. For the interchange of integration over the proper time and differentiation over $\Omega$ to be a valid operation, analytical properties of $F(\Omega)$ as the function of $\Omega$ are crucial. In this respect the presence of infrared regulator (e.g. nonzero temperature $T$) is important. Only at the final step one can remove it (e.g. to take zero temperature limit $\beta \to \infty$). In large temperature limit the function $D_\chi F(\Omega)$ greatly simplifies: it is given by power series in $\beta$ with known coefficients (see below). Closely related discussion can be found in \cite{garay}.

Let us make one more comment concerning abrupt switching limit. By way of example take window function (\ref{wf}) with the corresponding $D_\chi(s)$ given by (\ref{dt2}). From definition (\ref{d}) it is clear that small-$s$ expansion of $(D_\chi(s)-1)$ always starts from $s^2$-term. On the other hand in the limit $\lambda \to \infty$ the function $D_\chi(s)$ (\ref{dt2}) becomes $ D_\chi(s) = \left( 1 - {|s|}/{\tau_0} \right) \theta(\tau_0 - |s|) $
and the corresponding integral for the excitation probability
\be
P_{0\to 1} = {\bar g} \tau_0 \int\limits_{-\tau_0}^{\tau_0} ds \left( 1 - {|s|}/{\tau_0} \right) \> G^+(s) \exp(-i\Omega s)
\ee
becomes divergent. In particular, the limit $\tau_0 \lesssim \epsilon $ formally corresponds to Zeno regime:
$
P_{0\to 1} \sim {\tau_0^2} / {\epsilon^2}
$. To keep $\epsilon \neq 0$ is important (as was stressed in \cite{padm}) to get correct limit for excitation probability if $\tau_0 \to 0$ - switching detector for zero time must be equivalent to not switching it at all, so $P_{0\to 1}$ must go to zero in this limit.
This $p \sim \tau_0^2 \cdot G(0)$ scaling is to be compared with constant asymptotic for the ratio $p / \tau_m$ at large times (and hence, besides other things, exponential decays of excited states). The $\sim \tau^2$ behavior is crucial from quantum Zeno effect point of view \cite{zeno}, and corresponding inverse Zeno time is proportional to $\sqrt{G^+(0)}$.

We will continue the discussion using the language of macrodetectors, which is more suitable for us here. Each macrodetector consists of many microdetectors, which are taken as two level systems described by (\ref{pp24}), (\ref{pp214}). It is supposed that one is monitoring the relative ratio of excited to non-excited microdetectors in any given moment, but is not interested to follow individual record of every microdetector.

 First, it is instructive to consider scenario with two identical macrodetectors, one with all its microdetectors in the ground state (let us denote it as the detector A) and another one (the detector B) with all its microdetectors in excited state, put in contact with the field bath for some time. If this time is long enough in the sense of (\ref{par}), the detectors fully thermalize and all information about the initial state is totally erased. Speaking differently, there is exponentially small chance to learn which detector was at what initial state (excited or ground) observing their current states. Of course, the same is true for all other possible initial states. On the other hand, if the time period of the detectors being switched on is not long enough (for example, so that perturbative approximation is valid), the erasure is only partial. The magnitude of that is controlled by how many microdetectors changed their state due to interaction. Since for each macrodetector this quantity is perturbatively small (proportional to ${\bar g} \tau_m$) in weak coupling regime, it is instructive to consider the ratio of number of excited microdetectors in the macrodetector A to number of decayed microdetectors in the macrodetector B (both numbers are zero without interaction).
If one writes down the probability evolution expression using (\ref{jh}), (\ref{p12}) in the following form:
\be
p = {\cal P}_{0 \to 1} + p(\tau_0) \cdot e^{-\int\limits_{\tau_0}^{\infty} d\tau (C_+(\tau) + C_-(\tau))}
\label{i8}
\ee
where $\tau_0$ is some arbitrary chosen initial moment and $p(\tau_0)$ is probability that the microdetector is in its excited state at this moment,
the desired ratio is nothing but transition ratio
\be
\xi = \frac{{\cal P}_{0 \to 1} }{  {\cal P}_{1 \to 0}}
\label{tr5}
\ee
since $p(\tau_0)$ is zero for the macrodetector A and unity for the macrodetector B.
It is easy to see that the occupancy ratio $p /(1-p)$ reaches $\xi$ in equilibrium (given by $\xi = e^{-\beta \Omega}$ in case of full thermalization).

 Generally speaking, $\xi$ defined by (\ref{tr5}) is ${\bar g}$-dependent, but this dependence, as we already discussed, is encoded in normalization of $\chi(\tau)$. Quantities, independent on this normalization, are also ${\bar g}$-independent.
This is just the case in the leading order in perturbation theory, where
 it can be represented in the following way:
\be
{\xi} = 1 + \frac{\int d\tau \int d\tau' \> \chi(\tau) \chi(\tau')  \> e^{-i\Omega s} \> \Delta(\tau, \tau')}{\int d\tau \int d\tau' \> \chi(\tau) \chi(\tau')  \> e^{-i\Omega s} \> G^+(\tau', \tau)}
\label{comm}
\ee
where the presence of field commutator $\Delta(\tau, \tau') = \lan 0 | [\phi (x(\tau)), \phi (x(\tau'))] | 0 \rrr$ vanishing for classical $c$-number fields highlights quantum nature of ${\xi}$.

The leading perturbative term, according to (\ref{p8}) is given by
\be
{P}_{0\to 1} =  {\bar g } \> \tau_m \> D_\chi F(\Omega)
\label{r3}
\ee
It is easy to check that this coincides with (\ref{pert}), as it should be. The ratio reads
\be
{\xi} = \frac{D_\chi\> F_\beta(\Omega)}{D_\chi\> F_\beta(-\Omega)} =
1 - \frac{\Omega}{2\pi} \cdot \frac{1}{D_\chi\> F_\beta(-\Omega)}
\label{xi4}
\ee
where we took into account that  $D_\chi( - \partial^2/\partial \Omega^2 )\> \Omega = \Omega$.
In infinite measurement case $D_\chi \equiv 1$ and one immediately gets the standard thermal distribution ${\xi} = e^{-\beta \Omega}$ or
\be
{\bar\xi}_{mn} = \frac{P_{m\to n} }{P_{n\to m} } = e^{\beta (E_m - E_n) }
\ee
in multilevel case.

\section{Finite time effects}

\subsection{GZK horizon as Unruh-DeWitt detector excitation time}

Before we come to discussion of general structure of the operator $D_\chi$ and finite time corrections, let us make pedagogically instructive step by considering concrete perturbative example of the measurement with finite $\tau_m$ when the leading term $D_\chi = 1$ dominates. This is celebrated Greizen-Zatsepin-Kuzmin bound \cite{g,zk} on cosmic rays energy due to the presence of cosmic microwave background radiation. Despite the link between GZK bound and finite time quantum measurements\footnote{To the best of the author's knowledge this analogy has not been discussed in the literature before.} seems to bring no original result, it is of methodical interest. The physics of the bound follows from the fact that ultra-relativistic proton in outer space can be excited to higher hadron resonances by absorbing a thermal photon from the background, and the probability of this process increases with the proton energy and with the interaction time. On the other hand, dominant decay modes of these resonances are pionic, with corresponding significant energy loss. Of course, the scalar coupling considered by us here is a great simplification comparing with real photoproduction processes and no quantitative correspondence is to be expected. The key qualitative features however are identical. To describe this process in terms of (\ref{r3}) we are to take into account that for the detector (i.e. the proton with mass $m$ in our case) uniformly moving with the energy $E$ and speed $v$ the trajectory has the form $(\gamma \tau, 0, 0, v \gamma \tau)$ where $\gamma = E/(mc^2)$. We further assume that $\gamma \gg 1$, i.e. $v$ is close to 1. The detector excitation spectrum starts from some threshold parameter $E_{thr}$ of the order of the pion mass and goes up to infinity.
Plugging this world-line into (\ref{wi2}) and (\ref{r3}) one gets for the excitation probability
\be
 p = t_m  \sum\limits_n \frac{{\bar g}_{n}}{4\pi} \> \sum\limits_{l=1}^{\infty} \frac{1}{\gamma^2 \beta l} \> \exp\left(- \frac{\beta l \Omega_{n} }{2\gamma} \right)
\label{p81}
\ee
where ${\bar g}_{n} = g^2 | \lan \Delta_n | {\bf\mu}(0) | p \rrr |^2$, $\Omega_{n} = M({\Delta_n}) - m$. Notice that $t_m = l_m /v $ is physical time here, not proper time. In our model treatment the detector excitation spectrum consists of isolated $\Delta$-isobar like poles; of course, in reality it is  continuous above threshold, the fixed world-line approximation is also an over-simplification and the sum in (\ref{p81}) is to be replaced by the integral with integrand proportional to the corresponding photoproduction cross section. This would essentially give the expression obtained in original paper \cite{zk}. For not fast enough protons $\gamma k_B T \ll E_{thr}$ the excitation process in unit time has exponentially low probability and, correspondingly, exponentially large time is needed to get $p$ in the ballpark of unity. However when the regime $\gamma k_B T \lesssim E_{thr}$ is reached, the exponential suppression goes away and the relevant proton energy scale is given by $E_p^{crit} \sim (E_{thr} /4k_B T) \sim 10^{20}$ eV for $T = 3^\circ$, which corresponds to GZK horizon estimate $l_m$ in the ballpark of a few Mpc \footnote{Assuming the dimensionless ${\bar g}_{n}$ coupling of the order one, of course the horizon becomes proportionally larger for smaller couplings.}. It is remarkable that such a crude approximation provides correct order of magnitude estimates.

\subsection{Finite time thermometrics}

In the above example perturbative treatment was in line with the neglect of finite time corrections.
Let us now discuss general structure of such terms. At large measurement times one can expand $D_\chi$ as \cite{ya2}:
\be
D_\chi( -\partial^2/\partial \Omega^2 ) = 1 + \frac{1}{2\tau_{eff}^2} \frac{\partial^2}{\partial \Omega^2} + ...
\label{as2}
\ee
where we define positive quantity
\be
\frac{1}{\tau_{eff}^2} = \frac{\int d\bar\tau \> (\chi'(\bar\tau))^2 }{\int d\bar\tau \> (\chi(\bar\tau ))^2 }
\label{tau0}
\ee
having the meaning of effective interaction time. By no means should it coincide with the measurement time $\tau_m$, even in parametric sense. The instructive examples can be found in the Appendix B. For one-parametric Gaussian and Cauchy-Lorentz (\ref{chig}) shapes one indeed has
\be
\tau_m \sim \tau_{eff} \sim \tau_0
\ee
On the other hand for two-parametric hyperbolic tangential shape (\ref{wf}) one gets (in soft switching limit):
\be
\tau_{eff}^2 = \frac32 \> \tau_0 \cdot \delta \tau
\label{teff}
\ee
where $\delta \tau$ is switching time.
The scaling $\tau_{eff} \sim \tau_0^{1/2}$ remarkably demonstrates that the system in perturbative regime has a memory about its switching history even in the limit $\tau_0 \gg \delta \tau$ (but still small in order weak coupling approximation to be valid). Important conclusion is that universal character of $1/\tau_{eff}^2$ asymptotic (\ref{as2}) could well mean actual $1/\tau_m$ (and not naively expected $1/\tau_m^2 )$ dependence for the leading finite time correction.

Thus the leading correction is universal in the sense that it does not depend on the exact profile of $\chi(\tau)$ function but only on its first integral moment. Indeed, plugging the expansion (\ref{as2}) into (\ref{xi4}) one finds
\be
{\xi} = e^{-\beta \Omega} \left( 1 + \frac{\beta^2 \kappa(\beta \Omega)}{2\tau_{eff}^2}  \right)
\label{xi6}
\ee
where $ \kappa(x) = \mbox{cth} \frac{x}{2}  - \frac{2}{x}$.
In high temperature limit $\beta \Omega \ll 1$ one has $\kappa(x) \sim {x/6} $,
which means that at this order the distribution has effective temperature $\beta^* $:
\be
T^* = T\left(1+\frac{\hbar^2}{12 \tau_{eff}^2 (k_B T)^2}\right)
\label{limt}
\ee
This result was reported by the author earlier in \cite{ya2}. It coincides with the finite time temperature correction obtained in \cite{garay} using very similar formalism.

Speaking from a different prospective one can generally use the detectors in questions as thermometers in two ways. In the first algorithm one should wait until the macrodetector is thermalized (the initial state is not important in this case) and check the ratio of unexcited to excited microdetectors. This allows to compute the temperature (up to, perhaps, exponentially small corrections). The second strategy is to prepare macrodetector in special initial state (unexcited A and excited B), put it in contact with the bath for the time large enough the expansion (\ref{as2}) to be applicable, but small with respect to thermalization time, check the ratio (\ref{tr5}), and compute $\beta^*$ from (\ref{xi6}). The procedure returns higher values of the bath's temperature because of finite measurement time power  corrections.\footnote{Of course, an accuracy would also depend on the actual number of microdetectors in the macrodetector, we assume it to be large enough not to take that into account.}

Most thermometers people use in everyday life operate according to the first method and need quite some time to thermalize with a system being measured and indicate accurate result. The expression (\ref{limt}) corresponds to quantum uncertainty for the second measuring method (see \cite{correa} in this respect). In large temperature limit the correction is universal in a sense that it has no explicit $\Omega$-dependence (coding the thermometer microstructure, in a sense). Low temperature limit $\beta \to \infty$ is more tricky since the expansion (\ref{xi6}) takes $\beta/\tau_{eff} \ll 1$ as small parameter, according to the discussion above.

Making use of thermodynamic identity
\be
\frac{d}{d\beta} S(\beta) = -\beta \>\mbox{var}_\beta \left( H \right)
\ee
one conclude that the detector's entropy change is given by
\be
S^* - S = \frac{\hbar^2}{12 \tau_{eff}^2 (k_B T)^4 } \> \left(\lan E^2 \rrr - \lan E \rrr^2 \right)
\label{entropychange}
\ee
This result has some analogy with well known phenomenon of time-dependent coarse graining - having a given amount of Shannon information $I$ (for example, some coded signal), one can typically restore for finite time only coarse grained version of it (i.e. a signal containing less information $I^* < I$).

\subsection{Regimes of heating}

One can also pose another question - what is the softest possible mode for switching, i.e. optimal time profile for $\chi(\tau)$ which minimizes the finite-time corrections given fixed measurement time. In general case this is a complicated variational problem in the spirit of optimal control theory \cite{p}. However in large time limit it gets simple - according to (\ref{tau0}) one should minimize $\int d\tau (\chi'(\tau))^2 $ with fixed measurement time $\int d\tau (\chi(\tau))^2 $. Since $\tau_{eff}$ is independent on normalization of $\chi(\tau)$, without loss of generality the answer reads:
\be
 \chi_{opt} (\tau) = e^{-|\tau|/\tau_0}
\ee
Thus exponential profile is the softest, $\tau_m / \tau_{eff} = 1$ for it (this ratio is approximately 1.25 and 1.11 for Gaussian and Cauchy-Lorentz profiles, respectively).

Another interesting question is about possibility to reproduce thermal level distribution by some choice of the switching function. In other words, we are looking for $\chi(\tau)$ such that the resulting detector levels distribution is quasi-thermal at inverse temperature $\bar\beta$, despite the original bath is at zero temperature. One can see that this corresponds to the following condition
\be
D_\chi F_{\infty}(\Omega) = F_{\bar\beta}(\Omega)
\label{pj}
\ee
In principle, nothing guarantees the existence of regular solutions to (\ref{pj}). This however happens to be the case.
Expanding in Fourier components $\chi(\omega) = \int d\tau \chi(\tau) e^{i\omega \tau}$ with time-symmetric switching function, which corresponds to real $\chi(\omega)$, and taking into account that
\be
\frac{\partial^2 F_\infty(\omega + \Omega) }{ \partial \omega^2}  = \frac{\delta (\omega + \Omega)}{2\pi}
\ee
one obtains the following result
\be
\chi_{\bar\beta}(\tau) = \int d\omega \left(\tau_m \> \frac{\partial^2 F_{\bar\beta}(\omega)}{\partial \omega^2} \right)^{1 / 2}
e^{-i\omega\tau}
\ee
where the factor ${\tau_m}^{1/2}$ provides correct normalization. This "heating by switching" has much in common with heating by acceleration which is at the heart of Unruh effect. An interesting question about possibility to distinguish \cite{private} these "different types of heating" is addressed in \cite{padm5} (see also \cite{ut}).

\subsection{Switching off}

Let us look at the situation when the detector is already thermalized and being perturbatively switched off. To see what the corresponding dynamics is about, it is instructive to proceed with some concrete example of the switching off profile, which we take here as $\chi_-(\tau) = (1-\tanh(\lambda \tau) )/2$.
Since
\begin{equation}
\int\limits_{-\infty}^{\infty} d\tau \chi_-(\tau) (\chi_-(\tau - s) - \chi_-(\tau))
=  \frac{1}{2\lambda} \left(\lambda s(1-\coth ( \lambda s ) ) +  1 \right)
 \end{equation}

 one gets for the final probability
 \be
  p_f = p_i - p_i \int\limits_{-\infty}^{\infty} d\tau \> C_+(\tau) + (1-p_i) \int\limits_{-\infty}^{\infty} d\tau \> C_-(\tau)
  \label{pd}
 \ee
 the following answer at leading perturbative order:
\be
{p}_f = p_i + {\bar{g}} \left(\frac12-p_i\right) \cdot  \left[ I_\beta + I_p - \zeta^{(r)} \right]
\label{ddp2}
\ee
where
\be
I_\beta = \frac{1}{\lambda} \cdot \int\limits_0^\infty ds   \left(\lambda s(1-\coth ( \lambda s ) ) + 1 \right) \cdot \cos \Omega s \cdot  \left({\tilde G}_\beta^+(s) - {\tilde G}^+(s)\right)
 \label{i5}
\ee
\be
I_p = \frac{1}{\lambda} \cdot \int\limits_0^\infty ds   \left( 1 - \lambda s \> \coth ( \lambda s ) \right) \>  \cos \Omega s \cdot {\tilde G}^+(s)
\ee
where $\zeta^{(r)}$ is given by (\ref{z1}) and to get analytical result, the switching off start time $\tau_0$ is taken at minus infinity, of course, actual switching starts at the moment $\tau \sim -1/\lambda $. It is also worth noticing that due to the identity
\be
F(\Omega)(1-p_i) - F(-\Omega)p_i = 0
\ee
where $p_i$ is thermal distribution $p_i = 1/(1+e^{\beta \Omega})$, the right hand side of (\ref{pd}) is different from $p_i$ only if $d\chi(t) / dt \neq 0$. Physical meaning of that is clear - in stationary case initial thermal distribution stays intact for arbitrary large time.

The remarkable property of expression (\ref{ddp2}) is non-vanishing of the difference $p_f - p_i$  in the adiabatic limit $\delta \tau = 1/\lambda \to \infty$, where $\delta \tau$ is typical switching time. It seems, naively, that adiabatically slow switching should keep initial thermal state intact. Indeed, the probability of transition in unit time (given by the integral in (\ref{i5})) decays as $1/\delta\tau$ in large $\delta\tau$ limit. The probability however is an integral over time (this is reflected by the multiplier $1/\lambda = \delta \tau$ in front of the integral). Notice also that the cancellation of $\tau_s$-dependence at the leading order mentioned above does not take place here because initial value of switching function is not zero, $\chi_-(-\infty) = 1$.

As a result one gets
\be
\frac{{p}_f - p_i}{{\bar{g}} \left(1/2 - p_i \right)} =  \frac{1}{2\pi^2} \mbox{Ci}(|\Omega| \tau_s)
+ \frac{1}{2\beta^2 \Omega^2} \int\limits_0^\infty ds  \> s  \cos s \left[
{\left(\frac{\beta\Omega}{\pi s}\right)^2 - \mbox{csch}^2\frac{\pi s}{\beta \Omega }} \right]
\label{y4}
\ee
The second term in the rhs of (\ref{y4}) represents thermal contribution and vanishes in small temperature or large gap limit $\beta \Omega \to\infty$, while the term, proportional to integral cosine is the temperature-independent vacuum contribution (also vanishing in large gap limit). Its oscillatory behavior is a remarkable remnant of logarithmic divergency in the original problem.

The opposite limit, $\lambda \to \infty$ corresponds to abrupt switching and lies outside the domain of applicability of our formalism (see discussion above). Nevertheless it is instructive to notice that indeed $p_f \to p_i$ in this limit, as one would expect on general physical grounds.

\section{Finite time Landauer's principle}

We are now in the position to address the question posed at the beginning of this paper - what is the energy price of erasure done by external  force. The role of the latter is of course played by the switching function $\chi(t)$. As was discussed in the introduction, operationally erasure corresponds to some time-dependent actions like doing operations with the memory unit, removing potential barriers etc, or making contact between the memory unit and the bath and so on. Generally speaking, one can imagine situations when erasure requires {\it no} external work at all and also cases when erasure is done {\it by means} of the external work. Since finally this external work will be dissipated to the bath anyway, Landauer's bound, which is formulated in terms of this bath energy, should be applicable in both cases.

To proceed, let us rewrite the equation (\ref{l}) in more familiar form taking into account overall energy balance.
Inital state of the detector is described by the standard density matrix with some initial probability $p_i$. This detector can be seen as a pixel of the macro-detector, as described above. Using the fact that the detector in question has only two levels, we can represent the probability as $p_i=1/(1+e^{\beta_0 \Omega})$. Here the  "inverse temperature" $\beta_0$ can be positive (in case $0\le p_i < 1/2$) or negative (for $1/2 < p_i \le 1$) depending on the detector's initial state.

Energy balance equation has the following form:
\be
E^*_\phi + E^*_d = E_\phi + E_d + \delta E
\ee
where $E_d, E_\phi$ and $E^*_d, E^*_\phi$ are the initial and the final energies of the detector and the bath, respectively, while $\delta E$ is the work done by external force responsible for establishing bath-detector coupling.
For the probability to find the detector in excited state (\ref{o2}) at the moment $t$
we can, analogously, introduce $t$-dependent "inverse temperature" $
\beta^*  = \Omega^{-1}\log\left( {1} / {p(t) } - 1 \right)
$
which may vary from $-\infty$ to $\infty$.

The Landauer principle dictates
\be
\beta \Delta Q - \Delta S \ge 0
\ee
where $ \Delta Q = E^*_\phi - E_\phi\;\; ; \;\;
\Delta S = S_d - S_d^* $, or, in other words
\be
\beta \delta E \ge (\beta E^*_d  - S_d^*) - (\beta E_d  - S_d)
\label{ol1}
\ee
where change in the detector's energy is given by $
E^*_d - E_d = \Omega (p(t) - p_i)
$

The equation (\ref{ol1}) puts lower bound on the work done by external force till the moment $t$. The physical meaning of that is clear - increase of the detector's free energy (right hand side of (\ref{ol1})) cannot exceed the work done by external force (left hand side), otherwise one would be able to construct perpetuum mobile.

In terms of effective temperatures introduced above the condition (\ref{ol1}) takes the form:
\be
\beta \delta E \ge {\cal F}_\beta(\beta^*) - {\cal F}_\beta(\beta_0)
\label{ol13}
\ee
where the function ${\cal F}_\beta(z)$ is given by
\be
{\cal F}_\beta(z) = \frac{(\beta - z)\Omega}{1+e^{z\Omega}} - \log\left( 1 + e^{-z\Omega}\right)
\label{ol2}
\ee
Let us consider relevant limiting cases.

1. Thermalization. Full thermal equilibrium is characterized by $\beta^* = \beta$.  Since the function (\ref{ol2}) has its minimum at $z=\beta$, we conclude, that no nontrivial (i.e. positive) bound exists for external work. In other words, if the erasure is done by putting the detector in contact with thermal bath for unlimited time, Landauer's bound is always valid regardless initial state of the detector and there is no constraints on external work. Let us stress once again, that thermalization is not the "reset" operation.

2. On the other hand, suppose that the detector is initially in the thermal state, i.e. $\beta_0 = \beta$. Then for any $\beta^* \neq \beta$ one has positive bound for $\delta E$. Speaking differently, decoupling of the detector from the thermal bath cannot in general be done without external work. Notice that it is true not only for the case then the decoupled detector has effective temperature higher than the bath temperature (i.e. $\beta^* < \beta$) but also in the case then it is effectively colder (i.e. $\beta^* > \beta$). What matters is the balance of free energies and not energies themselves and in both cases these temperature difference can be used to extract useful work.

3. Let us come to a case when the detector is initially in its ground state, $p_i=0$, which corresponds to ${\cal F}_\beta(\infty) = 0$. One would say that the final levels full thermalization $\beta^* = \beta$ could be achieved without external work since the detector takes all required energy from the heat bath in this case. Crucial difference between erasure by thermalization and erasure by choosing the unique final state ("reset") can be seen here. In both cases the operation is irreversible, the information about initial state of the detector is completely lost (and this justifies the term "erasure"), but the entropy of the final state is zero in the latter case (and therefore never larger than the initial entropy), but the standard thermal entropy in the former one, and as such it can exceed the initial state entropy.

In the light of that it is seen that the point where the bound (\ref{ol1}) starts to be nontrivial corresponds to the final "effective inverse temperature" ${\bar\beta}^*$ such that
\be
{\cal F}({\bar\beta}^*) = 0
\label{lkj}
\ee
Sign of the unique solution to this equation coincide with the sign of $(\beta\Omega - 2\log2)$ and always ${\bar\beta}^* < \beta$.
 For $\beta^* <  {\bar\beta}^*$ one gets positive bound for $\delta E$. In other words, switching protocols $\chi(t)$ leading to $p(t)$  such that $\beta^*$ defined above is smaller than the solution to (\ref{lkj}) cannot be realized with zero external work.
Of course, for monotonous evolution of $p(t)$ the inverse temperature lowers from infinite initial value to $\beta$ and the $\beta^* <  \beta$ case cannot happens. In other words, the detector is never hotter than the bath. As we have shown, it could not be the case for finite interaction time.

In most cases of practical interest the energy gap of information bearing degrees of freedom - the detector in our case - is much larger than the thermal excitation energy, $\beta\Omega \gg 1$. Then with exponential accuracy ${\bar\beta}^*  = \beta - 1/\Omega$.
This defines the critical probability:
\be
p_{crit} = \frac{1}{1+ e^{\beta \Omega -1 }}
\label{ui}
\ee
In other words, this is the highest level of non-thermality allowed without external work penalty.

Comparison of (\ref{ui}) with (\ref{xi6}) gives estimate
\be
\tau_{eff}^{crit} = \beta \left( \frac{\kappa(\beta \Omega)}{2(e-1)}\right)^{1/2}
\ee
for the minimal time when erasure is possible for free. Notice $\tau_{eff}^{crit} \sim 1/T$ scaling in low temperature limit.

\section{Conclusion}

Main part of the present paper is devoted to study of finite time effects for Unruh-DeWitt detector. They are parameterized by time-dependent coupling $g\chi(\tau)$. We describe the detector's dynamics as Markov evolution (\ref{pp24}), (\ref{pp214}).
In principle, having fixed time profile $\chi(\tau)$, one is able to compute nonperturbatively the probability to find the detector excited at a given time in this framework. This is the input and new results obtained in the present paper with it are the following.
First, a new point is renormalization procedure which happens to be different in finite and infinite time cases. It is encoded in the expression (\ref{gb222}). We demonstrated that, as is typical for renormalization, a new parameter appears, having meaning of detector's recovery time $\tau_s$. It plays no role at the leading order of perturbation theory but can be important non-perturbatively.

Second, we have considered the case of coupling different from zero for finite time. The leading effect has linear scaling with this interaction time, and celebrated Greizen-Zatsepin-Kuzmin effect is used as good methodical example. We have systematically analysed the structure of next-to-leading terms, i.e. finite time corrections. Main elements of this analysis were reported in \cite{ya2} for perturbative setup (while in the present paper we explore both leading order results and the exact ones). It is found that in large-time limit they can be described in terms of the function/operator $D_\chi$ defined by (\ref{d}). There is non-perturbative exact relation (\ref{par}), expressing thermalization process through the same operator. In large-time limit the leading correction to perturbative answers has universal form (\ref{as2}). This brings finite-time corrections to temperature measurements. Other interesting finite time effects like heating by switching are touched.

Third, we considered another setup with perturbative adiabatic switching off the detector, which had been initially in equilibrium with the thermal bath. We found that such switching off leaves non-vanishing corrections to the detector's levels distribution. The reason for that is a compensation resulted from multiplying transition probability in unit time $\sim 1 / \delta \tau $ by total switching time $\delta \tau$.
The resulting correction (\ref{y4}) is sensitive to the detector's recovery time $\tau_s$.

One can think of time-dependent coupling as an eraser, while two-level Unruh-DeWitt detector - as an information bearing degree of freedom encoding one bit of (classical) information. Then this system has to obey the standard limits put by the Landauer's bound \cite{rl}.
Since the time-dependent coupling $g\chi(\tau)$ plays a role of external operating force in this case, the bound can be interpreted in terms of limits to external work by this force. For the detector in vacuum at zero temperature this limit has the simple form $\delta E \ge E_d^* - E_d$, i.e. external work must not be smaller than the difference between final and initial detector's energies (since if the detector is excited, it has happened as effect of external force). There is no information-theoretic background in this equation. It comes into play at nonzero temperature in the form of  corresponding entropies and this is the genuine content of Landauer's principle. In other words, we speak about external work which an eraser must do to perform erasure, and not about the energy dissipated into the surrounding non-information bearing degrees of freedom (despite finally everything goes to the bath anyway). This bound (\ref{ol13}) is nontrivial because in some cases erasure does not require external energy, while in other ones it does. It is important that while thermalization of the detector can be done without external work, it is generally not true for decoupling of the detector from the thermal bath. It is also interesting that there is the highest allowed level of non-thermality of the detector compatible with zero external work (\ref{lkj}), (\ref{ui}).

It is very interesting to test time-dependence patterns of this kind of physics using variable control parameters,
and work in this direction is going on intensively from both theoretical and experimental side \cite{diaman2,gam,dew,lutz2,ff,horo,vedral,jun,adami22,faist,diana}. Of special interest is application of these ideas to complex information processing systems and living objects, from cells to humans. Various aspects of that has already been addressed \cite{sch,brem,smith,mehta,goold1,england}.

\section*{Acknowledgments}

The author acknowledges partial support from Russian Foundation for Basic Research, grant RFBR-ofi-m N 14-22-03030.

\section*{Appendix A}

We demonstrate how to get the evolution equations (\ref{pp24}), (\ref{pp214}) from the initial definitions (\ref{si}), (\ref{p0}) in this Appendix A. The main steps from \cite{nitzan}
are being followed, with some modifications with respect to the original work which to our view make the exposition more clear.

The starting point is the time-dependent probability, defined in eq. (\ref{p0})
\be
{\cal P}_{0\to 0}(t) = \lan 0,\Phi | U_\chi^\dagger(t) |0\rrr \lan 0 | U_\chi(t) |0, \Phi \rrr
\label{p00}
\ee
Its time derivative can be rewritten as
\be
{\dot{\cal P}}_{0\to 0}(t) \equiv d {\cal P}_{0\to 0}(t) / dt  = \left( d K(t,\xi) / d\xi \right)_{\xi = 0}
\label{e01}
\ee
where
\be
\begin{split}
&K(t,\xi) = \lan 0,\Phi |  U^\dagger_\chi(t) e^{ig \xi \chi(t)\phi(t) [\Pi_0, \mu(t)]} U_\chi(t)  | 0, \Phi \rrr = \\
& = \lan 0,\Phi | {\mbox{T}}\exp \left[ ig \int_{C_{\pm}} d\tau  \> \chi(\tau)\phi(\tau){\tilde\mu}(\tau) \right] |0, \Phi \rrr = \\
& = \lan 0,\Phi | {\mbox{T}_\sigma}\exp \left[ ig \int\limits_{-\infty}^{\infty} d\sigma (d\tau(\sigma) / d \sigma)  \> \chi(\tau(\sigma))\phi(\tau(\sigma)){\tilde\mu}(\tau(\sigma)) \right] |0, \Phi \rrr
\label{derivative}
\end{split}
\ee
with $\Pi_0 = |0\rrr \lan 0|$ and $ \tilde\mu(\tau) = \mu(\tau) + \xi \delta(t-\tau)[\Pi_0, \mu(t)] $.

 The integral goes along closed Keldysh-type contour $C_{\pm}$ from minus infinity to $t$ and back to minus infinity. Notice that $K(t,0)=1$ due to normalization condition. Formally speaking, the ordering along $C_\pm$ is done with respect to the variable $\sigma$ (we denoted that in the last line in (\ref{derivative})), such that the contour $C_\pm$ is formed by the even function $\tau(\sigma)$, $\tau(0) = t$, $\tau(\sigma \to \pm \infty) \to -\infty$. Delta-function above is defined in such a way that $\int d\sigma (d \tau(\sigma) / d \sigma) \delta (t - \tau(\sigma)) = 1$. However in what follows we will continue to use notation $\int_{C_\pm} d\tau$ for brevity.

We are now in the position to apply cluster expansion to the above expression with respect to the field $\phi$. However care is needed here because the operators $\mu(\tau)$ are not commuting. To illustrate the point, it is instructive to examine the expansion of (\ref{derivative}) in $g$ up to the 4-th order. We get
\be
\begin{split}
 & K(t,\xi) = 1 - g^2 \int_{C_{\pm}} \chi(\tau_1) d\tau_1  \int_{C_{\pm}}^{\tau_1} \chi(\tau_2) d\tau_2 \> \lan 0 | {\tilde\mu}(\tau_1) {\tilde\mu}(\tau_2) | 0 \rrr \cdot G(\tau_1, \tau_2) + \\
 & + g^4 \int_{C_{\pm}} \chi(\tau_1) d\tau_1  \int_{C_{\pm}}^{\tau_1} \chi(\tau_2) d\tau_2 \int_{C_{\pm}}^{\tau_2} \chi(\tau_3) d\tau_3  \int_{C_{\pm}}^{\tau_3} \chi(\tau_4) d\tau_4 \times \\
& \times  \lan 0 | {\tilde\mu}(\tau_1) {\tilde\mu}(\tau_2) {\tilde\mu}(\tau_3) {\tilde\mu}(\tau_4) | 0 \rrr \cdot G(\tau_1, \tau_2, \tau_3, \tau_4) + {\cal O}(g^6)
 \end{split}
 \label{d2}
\ee
where $
G_{12} = G(\tau_1, \tau_2) = \lan \Phi | \phi(\tau_1) \phi (\tau_2) | \Phi \rrr
$,
$$
G(\tau_1, \tau_2, \tau_3, \tau_4) = \lan \Phi | \phi(\tau_1) \phi (\tau_2) \phi(\tau_3) \phi (\tau_4) | \Phi \rrr
$$
and so on. We also assume that $\lan \phi \rrr = 0$ and all vacuum averages of products with odd number of field operators vanish.

 Let us suppose, following \cite{nitzan}, that the bath is made of free fields. It corresponds to factorization of all higher order correlators $G(\tau_1, .. \tau_k)$ for $k>2$ into the sum of products of the Gaussian ones $G_{12}$. This approximation would, of course, be incorrect for strongly interacting fields forming the bath. It is important that keeping Gaussian term is enough to thermalize the detector, interacting with the bath for sufficiently large time. For the purposes of the present work this approximation (free field bath) plays no essential dynamical role.

 We can compare (\ref{d2}) with the corresponding expansion of Gaussian approximation to it we will use in what follows $K(t,\xi) \to K_2(t,\xi) $:
 \be
K_2(t,\xi) =
  \lan 0 | {\mbox{T}} \exp \left[ -g^2 \int_{C_{\pm}} \chi(\tau_1) d\tau_1  \int_{C_{\pm}}^{\tau_1} \chi(\tau_2) d\tau_2 \> {\tilde\mu}(\tau_1) {\tilde\mu}(\tau_2) \cdot G_{12} \right] |0 \rrr
  \label{e4}
 \ee
 It is worth noticing that $T$-ordering acts only on the operators $\mu(\tau)$ in the above expression, since vacuum average over the bath has already been taken. The expansion up to the 4-th order reads:

\be
\begin{split}
 & K_2(t,\xi) = 1 - g^2 \int_{C_{\pm}} \chi(\tau_1) d\tau_1  \int_{C_{\pm}}^{\tau_1} \chi(\tau_2) d\tau_2 \> \lan 0 | {\tilde\mu}(\tau_1) {\tilde\mu}(\tau_2) | 0 \rrr \cdot G_{1,2} + \\
 & + g^4 \int_{C_{\pm}} \chi(\tau_1) d\tau_1  \int_{C_{\pm}}^{\tau_1} \chi(\tau_2) d\tau_2 \int_{C_{\pm}}^{\tau_1} \chi(\tau_3) d\tau_3  \int_{C_{\pm}}^{\tau_3} \chi(\tau_4) d\tau_4 \times \\
  & \times \lan 0 |  {\mbox{T}} \left\{ {\tilde\mu}(\tau_1) {\tilde\mu}(\tau_2) {\tilde\mu}(\tau_3) {\tilde\mu}(\tau_4) \right\} | 0 \rrr \cdot G_{12} \cdot G_{34} + {\cal O}(g^6)
 \end{split}
 \label{d22}
\ee
Notice the differences at $g^4$ order between (\ref{d22}) and (\ref{d2}):  $\tau_1$ instead of $\tau_2$ as the upper limit in the third integral (and hence remaining $ {\mbox{T}} $-ordering) and $G_{12} \cdot G_{34} $ product instead of the sum $G_{12} \cdot G_{34} + G_{13} \cdot G_{24}  + G_{14} \cdot G_{23} $ (in case of Gaussian factorization for the field $\phi$). Thus from comparison of (\ref{d22}) and (\ref{d2}) one concludes that it is possible to approximate $K(t,\xi)$ by $K_2(t,\xi)$ provided overlapping  diagrams\footnote{By "overlapping " we mean here all diagrams where orderings in $\mu(\tau_i)$ and in $\phi(\tau_j)$ are different.} are suppressed with respect to non-overlapping ones. It is important to stress that both overlapping and non-overlapping diagrams are present at any given order in perturbation series in the coupling constant $g$. Parametrically, contributions from non-overlapping diagrams to the integrals in the right hand side of (\ref{d22}) scale as $(g^2 \tau_0)^n$, where $\tau_0$ is measurement time (encoded in the profile of $\chi(\tau)$). The situation is different for overlapping diagrams where some powers of $\tau_0$  are replaced by typical thermal time $\hbar / kT$ (discussion of related issues can be found in e.g. \cite{nvk,we3}). Therefore the use of $K_2(t,\xi)$
instead of exact $K(t,\xi)$ is justified if the temperature is large enough (or measurement is not too fast) to provide Markov property of the process, i.e. de-excitation/excitation of the detector at the time $t$ is correlated with a single act of its previous excitation/de-excitation at the time $\tau$ and has no memory about the rest (in  particular, when the previous (de)excitation took place). This condition is always valid for infinite time measurements discussed in \cite{nitzan}.

  There is another observation concerning the structure of (\ref{e4}). Since we are to take derivative over $\xi$ at the point $\xi = 0$, only the zeroth and the first powers in $\xi$ can be kept in the exponent and all other terms will vanish. In other words for the purposes of our derivation the product ${\tilde\mu}(\tau_1) {\tilde\mu}(\tau_2) $ can be rewritten as
  \be
  {\tilde\mu}(\tau_1) {\tilde\mu}(\tau_2) = {\mu}(\tau_1) {\mu}(\tau_2) + \xi (\tilde{a} \cdot {\bf 1} + \tilde{b} \cdot \Pi_0 ) + {\cal O}(\xi^2)
  \ee
  where
  \be
  \begin{split}
\tilde{a} + \tilde{b} = \delta(t - \tau_2) \lan 0 | \mu(\tau_1) [\Pi_0, \mu(\tau_2)]  | 0 \rrr +  \delta(t - \tau_1) \lan 0 | [\Pi_0, \mu(\tau_1)] \mu(\tau_2)  | 0 \rrr = \\
= \mu_0^2 (\delta(t - \tau_1) - \delta(t - \tau_2)) e^{- i\Omega(\tau_1 - \tau_2)};
\end{split}
\ee
\be
\begin{split}
\tilde{a} = \delta(t - \tau_2) \lan 1 | \mu(\tau_1) [\Pi_0, \mu(\tau_2)]  | 1 \rrr +  \delta(t - \tau_1) \lan 1 | [\Pi_0, \mu(\tau_1)] \mu(\tau_2)  | 1 \rrr =\\
= \mu_0^2 (\delta(t - \tau_2) - \delta(t - \tau_1)) e^{ i\Omega(\tau_1 - \tau_2)}
  \end{split}
  \ee
  and also $\lan 0 | \mu(0) |1\rrr = \lan 1 | \mu(0) |0\rrr = \mu_0$ with energy gap $\Omega = E_1 - E_0 > 0$. We have used the completeness condition for the detector's states.

Taking into account that
\be
\begin{split}
\int_{C_{\pm}} \chi(\tau_1) d\tau_1  \int_{C_{\pm}}^{\tau_1} \chi(\tau_2) d\tau_2 \> \delta(t-\tau_2) f(\tau_1, \tau_2) =  - \int\limits_{-\infty}^t d\tau f(\tau, t) \\
\int_{C_{\pm}} \chi(\tau_1) d\tau_1  \int_{C_{\pm}}^{\tau_1} \chi(\tau_2) d\tau_2 \> \delta(t-\tau_1) f(\tau_1, \tau_2) =  \int\limits_{-\infty}^t d\tau f(t, \tau)
\end{split}
\ee
the terms proportional to $\xi$ takes the form:
\be
\begin{split}
 \xi {\bar g} & \int\limits_{-\infty}^t d \tau \chi(t) \chi(\tau) \left[ \left( G(t,\tau) e^{i\Omega(t-\tau)} + G(\tau,t) e^{-i\Omega(t-\tau)} \right)\cdot  {\bf 1} \right. \\
 - & \left. 2( \vphantom{e^{i\Omega(t-\tau)}} G(t,\tau) + G(\tau,t)) \cos(\Omega(t-\tau))\cdot \Pi_0   ) \right]
\end{split}
\ee
where we denote ${\bar g} = g^2 \mu_0^2$ for brevity. Taking into account (\ref{e01}) and (\ref{e4}) one gets finally
 \be
\frac{d {\cal P}_{0\to 0}(t)}{dt} = - (C_+(t) + C_-(t) ) \cdot {\cal P}_{0\to 0}(t) + C_+(t)
\label{pp2}
\ee
and
\be
\frac{d {\cal P}_{0\to 1}(t)}{dt} = - (C_+(t) + C_-(t) ) \cdot {\cal P}_{0\to 1}(t) + C_-(t)
\label{pp21}
\ee
where $C_+(t)$ is (un)renormalized coefficient function $C(t, \Omega) $ given by
\be
 C(t, \Omega)
={\bar g}\int\limits_{-\infty}^t d \tau \chi(t) \chi(\tau)  \left( G(t,\tau) e^{i\Omega (t-\tau)} + G(\tau,t) e^{-i\Omega (t-\tau)}\right)
 \label{gb}
 \ee
and $C_-(t)$ stays for $C(t,-\Omega)$. These expressions form the basis of our discussion in the present paper. Let us stress again that they are valid under two basic set of assumptions: free field (Gaussian) bath and sufficiently large (with respect to the thermal time) measurement duration. Under these dynamical assumptions the detector's evolution is Markovian and follows (\ref{pp2}), (\ref{pp21}).

\section*{Appendix B}

In this Appendix B we collect relevant formulas for concrete switching profiles.
We confine our attention to three shapes: Gaussian shape and Cauchy-Lorentz shape
\be
\chi_G(\tau) = e^{-\frac{(\tau - \bar\tau)^2}{\tau_0^2}} \;\;\; ; \;\;\; \chi_L(\tau) = \frac{\tau_0^2}{(\tau - \bar\tau)^2 + \tau_0^2}
\label{chig}
\ee
and, as more complex example, two-parametric profile:
\be
\chi_\theta(\tau) = \frac{\tanh(\lambda(\tau - \tau_1)) + \tanh(\lambda(\tau_2 - \tau))}{2\tanh (\lambda \tau_0 /2 )}
\label{wf}
 \ee
In the latter case $ \tau_{1,2}$ define the moments of proper time when the detector is switched on and off, respectively, and we assume $\tau_0 = \tau_2 - \tau_1 > 0$. The parameter $\lambda = 1/\delta \tau$ is the inverse switching time. It is usually reasonable to assume "large window" approximation $z=\lambda \tau_0 \gg 1$. In $\lambda\to\infty$ limit (abrupt switching) one has obviously
$
\chi_\theta(\tau) \to \theta (\tau - \tau_1) - \theta(\tau - \tau_2)
$. All above profile functions are normalized to unity.
In all these cases the operator $D_\chi (s)$ defined in (\ref{d}) can be explicitly computed, the results read:
\be
D_{\chi_G}(s) = e^{- s^2 / 2\tau_0^2} \;\;\; ; \;\;\;  D_{\chi_L}(s) =  \frac{1}{1 + (s / 2 \tau_0)^2}
\label{dl2}
\ee
and
\be
D_{\chi_\theta}(s) = \frac12 \frac{\mbox{sh} z}{\mbox{sh} \lambda s } \frac{1 }{z\mbox{cth} z -1} \left[ \frac{z - \lambda s}{\mbox{sh}(z- \lambda s)} - \frac{z + \lambda s}{\mbox{sh}(z+ \lambda s)}\right]
\label{dt2}
\ee
 In large temperature limit these expressions should be understood as asymptotic expansions in $1/\tau_0$. The leading term, with the substitution $s \to i \partial / \partial \Omega$  has the following form:
\be
D_\chi( \partial^2/\partial \Omega^2 ) = 1 + \frac{1}{2\tau_{eff}^2} \frac{\partial^2}{\partial \Omega^2} + ...
\label{asrr}
\ee
where $ \frac{1}{\tau_{eff}^2} = \frac{\int d\bar\tau \> (\chi'(\bar\tau))^2 }{\int d\bar\tau \> (\chi(\bar\tau ))^2 } $ has the meaning of effective interaction time. The choice of coefficients in (\ref{dl2}) provides $\tau_{eff} = \tau_0$ for Gaussian and Lorentz shapes. The case of $\chi_\theta (\tau)$ is more tricky. In the soft switching limit (small $\lambda$) one gets
\be
\tau_{eff}^2 = \frac{3\tau_0}{2\lambda}
\ee
The scaling $\tau_{eff} \sim \tau_0^{1/2}$ remarkably demonstrates that the system has a memory about its switching history even for $\tau_0 \gg \delta \tau$. Important conclusion is that universal character of $1/\tau_{eff}^2$ asymptotic (\ref{asrr}) could well mean actual $1/\tau_0$ (and not $1/\tau_0^2 )$ dependence for the leading finite time correction.


\end{document}